\documentstyle[11pt]{article}

\begin{document}
\topmargin 0pt
\oddsidemargin 5mm
\setcounter{page}{1}
\begin{titlepage}
\hfill Preprint YERPHI-1530(30)-99
\vspace{2cm}
\begin{center}

{\bf Doubly Charged Fermions Bound States,Vector Bosons Bound States }\\
\vspace{5mm}
{\large R.A. Alanakyan}\\
\vspace{5mm}
{\em Theoretical Physics Department,
Yerevan Physics Institute,
Alikhanian Brothers St.2,

 Yerevan 375036, Armenia\\}
 {E-mail: alanak @ lx2.yerphi.am\\}
\end{center}

\vspace{5mm}
\centerline{{\bf{Abstract}}}
We consider positronium-like bound states of doubly
 charged fermions.We consider also  $P(CP)$-parity
violation in positronium like system (including quarkonium and lepton
antilepton bound states and mesoatoms) which take
place due to $Z^0$-bosons or Higgs bosons exchange.
By model independent way we consider also
 vector bosons bound states in the Coulomb and Higgs potential
and have shown that in the presence of the Higgs potential
effective potential become less singular and in this case fall
down on the center is absent.
Vector bosons energy levels in the magnetic field are considered .
Considered magnetism of electron gas in the finite volume.

\vspace{4mm}
\vfill
\centerline{{\bf{Yerevan Physics Institute}}}
\centerline{{\bf{Yerevan 1999}}}

\end{titlepage}

{\bf Doubly charged fermions bound states}

It is known that left-right supersymmetric models
 predict the existence of doubly charged fermions
superpartners of  doubly charged Higgs bosons in the
left-right supersymmetric models ( for left-right supersymmetric model see
[1],\cite{DMM} and references therein, for pair production of doubly charged fermions
  in $e^+e^-$ colliders and its subsequent decays see \cite{DMM} \cite{RZ}
  and references therein).

In this paper we consider  bound states which consist  of
 doubly (and singly) charged Higgs bosons (for triplet Higgs bosons see \cite{M1} and
 \cite{GGMKO}) and its superpartners
$\delta^{++}_{L}\delta^{--}_{L}$,
$\delta^{+}_{L}\delta^{-}_{L}$ ,$\delta^{++}_{L}\delta^{-}_{L}$ ,
$\tilde{\delta^{++}}\tilde{\delta^{--}} $,
$\tilde{\delta^+}\tilde{\delta^-} $
,"charginium" (i.e. bound state which consist of
$\tilde{\chi^+}\tilde{\chi^-}$, for review of supersymmetric
theories see e.g. \cite{HK},\cite{GH} and references therein).

This bound state are positronium-like,
and for its binding energy we obtain from appropriate
formula for positronium levels (For positronium levels see e.g. \cite{BLP})
by substitution $\alpha \rightarrow \alpha Q^2 $ the following result:
\begin{equation}
\label{A1}
\Delta E_{n,l}=-\frac{\alpha^2Q^4m_H}{4n^2}-\frac{\alpha^4Q^8m_H}{2n^3}(\frac{1}{2l+1}-
\frac{11}{32n})
\end{equation}
where $m_H$ is the  mass of $\delta^{--}_{L,R}$-bosons or its superpartners.

As we show below however the effect of vacuum polarization
which is small in case of positronium in case
of heavy fermions is larger than the correction of order
$\frac{v^2}{c^2}$
described by last term in (1) (For vacuum polarization  see e.g. \cite{BLP}).

We consider also P,CP-violation in heavy fermion bound states,
in $l^-_il^+_j$ bound states and in mesoatoms
(for $\mu^+e^-$ atoms and mesoatoms see \cite{MM1} ,\cite{K} respectively).

 We would like to stress that the
width of the $\delta^{--}_{L,R}$-bosons, decays
$\Gamma(\delta^{--}_{L,R}\rightarrow l^-l^-
)=\frac{h^2}{8\pi}m_H\ll E_{n,l}$
at sufficiently small Yucawa coupling $h$ and system may be treated as the bound state.
The width of the doubly charged fermions decays
$\tilde{\delta^{--}_{L,R}}\rightarrow \tilde{l^-}l^-$
is the same order( for doubly charged fermions decays see \cite{DMM})
The main contribution to the width of bound state formed from
triplet Higgs bosons  or its superpartner comes predominantly  from its
own decays into leptons.
The bound states with quantum numbers $J^{PC}=1^{--}$
may be produced at $e^+e^-$ through
virtual $\gamma, Z^0$-bosons exchanges in resonance
\footnote{Bound state which consist of doubly
charged Higgs bosons at $L=0$ have quantum numbers $J^{PC}=0^{++}$
and can not be produced at $e^+e^-$-collisions in resonance, however
 in resonance may be produced P-wave bound state
of doubly charged Higgs bosons.Its decay width into $l^+l^-$ may obtained from
the width  of squarkonium into  $l^+l^-$,in squarkonium decay
into leptons also must be taken into account the effect of photon and Z-boson exchange
 (for squarkoniums, its production and decays see e.g.
\cite{GH},\cite{HK} and references therein).
We have $\Gamma(\delta^{++}_{L,R}\delta^{--}_{L,R}
\rightarrow l^+l^-)=\frac{\alpha^2Q^2|R'(0)|^2}{2m_H^4}
((1+Q^{-1}g_Vg^e_V)^2+Q^{-2}(g_Ag_A^e)^2))=\frac{\alpha^7Q^{12}m_H}{32^9}
((1+Q^{-1}g_Vg^e_V)^2+Q^{-2}(g_Ag_A^e)^2))$}.

At $m_H>>m_Z$ the width of the decay  of bound state of doubly
charged  fermions has the following form:
\begin{equation}
\label{2}
\Gamma(\tilde{\delta^{++}_{L,R}}\tilde{\delta^{--}_{L,R}}
\rightarrow l^+l^-)=\frac{\alpha^5Q^8m_H}{6n^3}
((1+Q^{-1}g_Vg^e_V)^2+Q^{-2}(g_Ag_A^e)^2))
\end{equation}
where $m_H$ is the mass of the doubly charged fermion,
$g_V,g_A(g_V^e,g_A^e)$-are vector and pseudovector
couplings of doubly charged fermion (leptons) to $Z^0$-bosons (we parametrize interaction of
Z-bosons with fermions in the form:$L=e \bar{f}\hat{Z}(g_V+g_A\gamma_5)f$).
From decay widths of orto- and para-positronium it is easy to obtain that:
the widhts of the decay  into $2\gamma$,$3\gamma$ is
enhanced in comparison with
para- and ortopositronium decay  widths ( into $2\gamma,3\gamma$)
 in $Q^{10},Q^{12}$ times respectively.

In resonance the Breit-Wigner cross section of the
$1^{--}$ bound state production  has the following form:
\begin{equation}
\label{3}
\sigma=\frac{12\pi}{4m_H^2}\frac{\Gamma(\tilde
{\delta^{++}_{L,R}}\tilde{\delta^{--}_{L,R}}
\rightarrow l^+l^-)}{\Gamma_{total}}
\end{equation}
As seen from (44) the width of the doubly charged
particles decay into leptons is comparable
with decay width of the ordinary quarkonium into leptons:
\begin{equation}
\label{4}
\frac{\Gamma(\tilde{\delta^{++}_{L,R}}\tilde{\delta^{--}_{L,R}}
\rightarrow l^+l^-)}{\Gamma(\bar{q} q \rightarrow l^+l^-)}\approx\frac{Q^8\alpha^3}{3Q^2_q\alpha^3_s}\approx
0.366 (at Q_q=-\frac{1}{3}),0.092 (at Q_q=\frac{2}{3})
\end{equation}
and with decay width of the squarkonium (quarkonium) into photons:
\begin{equation}
\label{5}
\frac{\Gamma(\tilde{\delta^{++}_{L,R}}\tilde{\delta^{--}_{L,R}}
\rightarrow \gamma\gamma)}{\Gamma(\bar{q} q \rightarrow l^+l^-)}\approx
\frac{Q^{10}\alpha^3}{3Q^4_q\alpha^3_s}\approx
13.184 (at Q_q=-\frac{1}{3}),0.824 (at Q_q=\frac{2}{3})
\end{equation}
Thus, in $\gamma\gamma$-collision \cite{G} doubly charged particles
bound states may be produced more copiously than
quarkonium with same mass.
If beam resolution is much larger than the total width of the bound state
the cross section (3) in $\sim \frac{\Gamma}{\Delta}$ times will be
smaller.
For exmaple, at $\Delta({\sqrt{s}})=10^{-3}\sqrt{s}
>>\Gamma(\delta^{--}_{L,R}\rightarrow l^-l^-)$ we obtain about
$52*Q^8$ events per year at $m_H=500 GeV$
and yearly luminosity $L=1000 fb^{-1}$.
The main decay  modes  are decays into
$l^+l^-,\bar{\nu}\nu,q\bar{q},2\gamma,3\gamma, H^0\gamma, Z^0\gamma$
 etc.

All above analys is also true for $\tau^+\tau^-$
(firstly decays of  $\tau^+\tau^-$,$\mu^+\mu^-$
bound states has been considered in \cite{JM}, however in this ref.has not been
considered its production in resonance)
bound states which also may be produced in $e^+e^-$ collisions in resonance and also in
$\gamma\gamma$-collisions.
$\mu^+\mu^-$ can also produced in resonance, and at $L=1 fb^{-1}$
we obtain 3886*400 events per year.
For instance at $\Delta({\sqrt{s}})=10^{-3}\sqrt{s}
>>\Gamma$ we obtain in accordance with above made conclusions
$3886$ $\tau^+\tau^-$ bound states  per year  at $L=1 fb^{-1}$.
$\mu^+\mu^-$ can also produced in resonance, and at $L=1 fb^{-1}$
we obtain 3886*400 events per year
(For $\mu^+\mu^-$ atoms see \cite{JM}) .Also
$\mu^+\mu^-$ atoms may be produced as a result of positron collision with electrons of target.
In this case depend on $n$-numbers of electrons in $cm^{-3}$ and width of the target
we obtain the
enhancement on many order in comparison with $\mu^+\mu^-$ atoms production
in $e^+$ and $e^-$ beams collision in c.m.system.

It must be noted that levels shift from
vacuum polarization in $\tau^+\tau^-$ is also essential.

In case of tau leptons bound state the main contribution in energy
shift from vacuum polarization come from electron loops
corrections.In case of doubly charged fermions bound states
contributes besides electrons also muons and even tau lepton and
pions loops (i.e. all light flavors: $a_B<<\frac{1}{m}$).

For vector doubly charged bosons (see \cite{F1}) this consideration is also true
(see "Vector bosons bound states" below).

{\bf Vacuum Polarization in  doubly charged fermions bound states}

The modification of Coulomb potential between fermions from corrections to
photon propagator has the following form (Uling potential):
\begin{equation}
\label{6}
V(r)=-\frac{2\alpha^2Q^2}{3\pi}\frac{1}{r}\int^{\infty}_1\exp(-2mrx)F(x)dx
\end{equation}

where $F(x)=(1+\frac{1}{2x^2})\frac{\sqrt{x^2-1}}{x^2}$.The
corrections to the energy levels may be calculated by perturbation
theory.
E.g. for ground state ($n=1,l=0$) energy level shift we have:

\begin{equation}
\label{7}
\Delta E_{1,0}=-\frac{\alpha^3Q^4m_H}{3\pi}\int^{\infty}_1\frac{1}{(m_exa_B+1)^2}F(x)dx
\end{equation}

For ($n=2,l=0$,$n=2,l=1$) energy levels shifts we have :
\begin{equation}
\label{8}
\Delta E_{2,1}=-\frac{\alpha^3Q^4m_H}{12\pi}\int^{\infty}_1\frac{(8m^2_ea_B^2x^2+1)}{(2m_exa_B+1)^4}F(x)dx
\end{equation}

\begin{equation}
\label{9}
\Delta E_{2,1}=-\frac{\alpha^3Q^4m_H}{72\pi}\int^{\infty}_1\frac{1}{(2m_exa_B+1)^4}F(x)dx
\end{equation}

In case of heavy fermions (doubly charged fermions bound
states,$\tau^+\tau^-$,...)$a_B<<\frac{1}{m_e}$ and it
is convenient instead  this general formulas  to use asymptotic
simplified formulas obtained from Uling potential at $m_er<<1$.
At $m_er<<1$ the contribution to potential from vacuum polarization is following \cite{AB}:
\begin{equation}
\label{10}
V(r)=\frac{2\alpha^2Q^2}{3\pi}\frac{1}{r}(ln(mr)+C+\frac{5}{6})
\end{equation}
where $C=0.577$ is Eiler constant.
In positronium vice versa $a_B>>\frac{1}{m_e}$ and Uling potential
has the behaviour \cite{BLP}: $\sim \frac{\exp{-2mr}}{r^{2.5}}$ which give
correction of order $\Delta E\sim-\alpha^5m$.
In our case we obtain  $\Delta E\sim-\alpha^3m_H ln(m_ea_B)$  and besides electron another
 light fermions ($a_B=\frac{2}{m_H\alpha Q^2}<<\frac{1}{m_e}$) contributing in
loop.

In $\tau^+\tau^-$ bound states only electrons give essential
contribution.In doubly charged fermions bound states case
contribution of muons approximately in  $\sim 3$ times smaller than electrons
contribution.In case of tau-leptons bound state the muons loop  contribution  is negligible.
 Also may be essential $\pi^+\pi^-$
loops because $a_B<<\frac{1}{m_{\pi}}$, resonance contribution
(e.g. $\rho$-mesons contribution etc.)As pointed in \cite{BLP} the
hadronic contributions in vacuum polarization is negligible in
positronium.

Using asymptotic behaviour of the Uling potential
e.g. for $n=1,2,l=0$ and  $n=2,l=1$  we obtain:
\begin{equation}
\label{11}
\Delta E_{1,0}=\frac{\alpha^3Q^4m_H}{3\pi}
(ln(\frac{m_ea_B}{2})+\frac{11}{6})
\end{equation}
\begin{equation}
\label{12}
\Delta E_{2,0}=\frac{\alpha^3Q^4m_H}{12\pi}
(ln(m_ea_B)+\frac{7}{3})
\end{equation}
\begin{equation}
\label{13}
\Delta E_{2,1}=\frac{\alpha^3Q^4m_H}{12\pi}
(ln(m_ea_B)+\frac{8}{3})
\end{equation}

Thus the difference between $n=2,l=0$ and $n=2,l=1$ levels is much
larger than in positronium.

In system  with $Q=1$  beginning from tau-leptons this
contribution is essentially smaller than vacuum polarization
effect however it become essential and comparable with vacuum polarization
effect at large charges $Q=2$.

Our numerical results is following: for $\tau$-lepton we have $\Delta E_{1,0}\approx
10^{-3},\Delta E_{2,1}\approx  10^{-4}$,
for $m_H\sim 500 GeV$ we have $\Delta E_{1,0}\approx 0.055E_{1,0},0.01E_{1,0}$ for $Q=1,2$
respectively,for $m_H\sim 500 GeV$ we have $\Delta E_{2,1}\approx 0.005E_{1,0},0.004E_{1,0}$
for $Q=1,2$ respectively.

Our results are applicable also to bound states of heavy quarks (several hundred GeV).
In heavy quarkoniums $a_B=\frac{2}{m\frac{4}{3}\alpha_s}<<1/300 MeV$ however as we see below
quarkonium cannot be considered as Coulomb-like system.
In this case as known  vacuum polarization comes from light quarks and gluons loops.We obtain
 $\Delta E_{1,0}\sim N\alpha_s^3ln(\frac{\mu}{m\alpha_s})\approx 1.2E_{1,0}\sim m\alpha_s^2$
(N-number of flavours, $\mu\sim 200 MeV$)
which mean that vacuum polarization cannot
considered as perturbation and consequently heavy quarkonium by this reason
cannot be considered as Coulomb-like system.Must be solved Dirac equations
numerically with potential $-4/3\alpha_s(r)/r$,
(where $\alpha_s(r)=\frac{\alpha_s(r_0)}{1-b\alpha_s(r_0)ln(\frac{r}{r_0})}$).

At small distances in potential must be included also repulsive
term $\alpha Q^2(\frac{(\vec{A})^2}{2m}\sim r^{-4}$
($\alpha_s(\frac{\vec{A}})^2{2m}\sim r^{-4}$ in case of QCD) which can not be considered
as perturbation because appropriate integral is divergent at small $r$.

(for Coulomb-like potential in quarkonium  see e.g. \cite{AN} and ref.therein).
Vacuum polarization in quarkoniums as perturbation has been considered in
\cite{PY}.

It must be noted in quarkonium or squarkonium or above mentioned
bound states of doubly charged fermions also give contribution
 to the potential from $Z$-bosons and Higgs bosons
exchanges(see below).This corrections may be especially large
for s-channel exchanges if mass of particle in the s-channel is
equal to the mass of the bound state.E.g. for s-channel exchange
case via scalar and vector bosons we have:
\begin{equation}
\label{14}
\delta E=h_s^2\frac{1}{s-m^2_H+i\Gamma
m_H}|\psi(0)|^2-
\alpha\frac{(g^2_V+g^2_A\vec{\sigma_1}\vec{\sigma_2})}{s-m^2_V+i\Gamma_Vm_V}|\psi(0)|^2
\end{equation}
We see that this corrections give contribution also in the width
of the bound state.

{\bf $CP$-Parity violation.}

We use the following parametrization of the scalar (pseudoscalar) particles
interaction with fermions:
\begin{equation}
\label{14}
L=\bar{f}(b_s+ib_P\gamma_5)f
\end{equation}

The contribution of scalar particles into CP-violating potential take the form(
for CP-violation in hydrogen see \cite{Kh} and also references therein, in positronium CP-violation has been cionsidered in
\cite{CHR},\cite{CKL} part of decays considered below has been studied in \cite{CHR}):
\begin{equation}
\label{15}
V(r,\vec{p})=b_Sb_P\frac{1}{2m}\vec{\Delta}\vec{n}H'(r)
\end{equation}
where at tree levels $H(r)=\frac{\exp{-m_Sr}}{r}$,
$\vec{\Delta}=\vec{\sigma_1}-\vec{\sigma_2}$,$\vec{n}=\frac{\vec{r}}{r}$.
This formula is true only at $v<<1$, in general it is necessary to
replace in the last formula $i\vec{n}\rightarrow(p_f-p_i)$ where $p_f,p_i$are momentums of the
initial and finite states.For potential $V=\vec{\Delta}\vec{n}F(r)$ we obtain
\footnote {The wave functions and its derivatives may be considered as
constants only if V(r) give essential contribution at $r<<a_B$}:
\begin{equation}
\label{16}
<1^{--}|V(r)|1^{++}>=\int \psi_P'(r)\psi_S(r)F(r)d^3r\approx \psi_P'(0)\psi_S(0)F(r)d^3r
\end{equation}
Also must be taken into account electric dipole moment of doubly charged fermions,
tau-leptons or quarks:
\begin{equation}
\label{A200}
L=A\bar{f}\sigma_{ab}\gamma_5fF_{ab}
\end{equation}
which give the following contribution  in CP-violating potential:
\begin{equation}
\label{17}
V(r)=eA(\vec{\Delta} \vec{n})\frac{1}{r^2}
\end{equation}
Due to  $CP$-parity violating interaction () may mixed states with different
$CP$-parity. For example can mixed $0^{-+}$ and $0^{++}$,
$1^{+-}$ and $1^{--}$ bound states.
Its lead to enhancement of some rare radiative decays.Besides may take place
some decays which are forbidden by  T-parity conservation.

In particular decays $0^{-+}\rightarrow1^{--}\gamma$,
$1^{--}\rightarrow0^{-+}\gamma$,$1^{--}\rightarrow0^{-+}\gamma$ are supressed as
$M1$ transition however as in hydrogen atom the small mixture of opposite parity
states ($1^{--}-1^{+-}$,$0^{++}-0^{+-}$ mixing may enhanced this decays, because
,$1^{+-}\rightarrow0^{-+}\gamma$
,$0^{++}\rightarrow1 ^{--}\gamma$  decays take place as $E1$ transition.

Also, Ora-Pauwell decay of orthopositronium into 3 photons (gluons) may be
enhanced in the range of soft photons (gluons),
because due to CP-parity violation $1^{--}$ is mixed with $1^{+-}$ bound
state which also may decay into three photons (for positronium and bottomonium
decays into 3 photons , 2 photons+gluon,3 gluons see \cite{N} and references
therein).

The differential width into 3 photons decays of the ${}^3 S_1$ state (which is
${}^3 S_1$ state with small mixture of ${}^1 P_1$ state) takes the form:

\begin{equation}
\label{18}
\frac{d\Gamma({}^3 S_1'\rightarrow3\gamma)}{d\omega}
\approx a^2_{T}\frac{d\Gamma({}^1 P_1\rightarrow3\gamma)}{d\omega}+
\frac{d\Gamma({}^3 S_1 \rightarrow3\gamma)}{d\omega}
\end{equation}
where
$\frac{d\Gamma({}^1 P_1\rightarrow3\gamma)}{d\omega}\sim\frac{1}{\omega}$,
whereas $\frac{d\Gamma({}^3 S\rightarrow3\gamma)}{d\omega}\sim\omega log
\frac{m}{\omega}$,
$\omega$ is photon energy, $a_T$ is the measure of the mixing  between two states with different
 T-parity:
\begin{equation}
\label{19}
|{}^3 S_1'>=|{}^3 S_1>+a_T|{}^1 P_1>
\end{equation}
\begin{equation}
\label{A200}
a_T=\frac{<{}^3S_1|V|{}^1P_1>}{E({}^3 S_1)-E({}^1 P_1)}
\end{equation}
Thus at small $\omega$ we have peak instead usual Ora-Pauell
formula. Analogous effect take place in quarkonium (decays into 3
photons,one photon+2 gluons, 3 gluons).
Analogously for above mentioned radiative correction for mixed ${}^3S_1'$ decays we
have:
\begin{equation}
\label{20}
\Gamma({}^3 S_1'\rightarrow {}^1 S_0 \gamma)
\approx
a^2_T\Gamma({}^1 P_1\rightarrow {}^1 S_0 \gamma)+
\Gamma({}^3 S_1\rightarrow {}^1 S_0 \gamma)
\end{equation}
In bottomonium the wave functions is not Coulomb like, however the
behaviour of the
matrix elements  are expressed through $R(0),R'(0)...$ which may be expressed
through decay widths of the bottomonium (e.g.
$\Gamma(0^{-+}\rightarrow2g)\sim|R_S(0)|^2,
\Gamma(0^{++}\rightarrow3g)\sim|R'(0)|^2$,
see \cite{N} and references therein).

{\bf $P$-Parity violation.}

In heavy fermion bound states become large P- and CP-violation effects
(because $Gm^2_H$ is large).The effect of P-violation in heavy fermion bound
states is analogous to the P-violation in positronium which was calculated in  \cite{MM1}.
Thus we can use  formula (B4) of the  \cite{MM1} for mixing coefficient between
 $1^{--}$ and $1^{++}$ bound states where has been made the following replacement(we take
into account that in heavy fermions bound states s-channel contribution is supressed as $\frac{m^2_Z}{m^2_H}$):
\begin{equation}
\label{21}
\alpha \rightarrow \alpha Q^2,g_V^e \rightarrow g_V, m_e\rightarrow m_H
\end{equation}

Using  the result for $E(2^3 S_1)-E(2^3P_1)$ obtained above
(effect of vacuum polarization and contribution from $v^2/c^2$
corrections described by formula (1))
(B4) of the  \cite{MM1} with previous replacements
we have for  $2^3 S_1-2^3P_1$ mixing:
\begin{equation}
\label{22}
a_P=-1.66*10^{-2}(\frac{m_H}{500 GeV})^2(g_V/g_V^e)
(\frac{1}{24}m\alpha^2Q^8/(E(2^3S_1)-E(2^3P_1)))
\end{equation}

Thus we see that although there is large enhancement in
comparison with positronium, the effect of $P,CP$-violation in
doubly charged fermions bound states case is again small for observation.
Due to  $P$-parity violation may mixed states with different $P$-parity.For example
can mixed
$1^{--}$ and $1^{++}$ bound states.It lead to enhancement of some rare
radiative decays and appearance of the new channels which has been forbidden
by $P$-parity conservation\footnote{most of this decays has been considered in \cite{MM1}}.

In particular decays $0^{-+}\rightarrow1^{--}\gamma$,
$1^{--}\rightarrow 0^{-+}\gamma$, are suppressed as
$M1$ transition however as in hydrogen atom the small mixture of opposite parity
states ($1^{--}-1^{++}$, mixing) may enhanced this decays, because
,$1^{+-}\rightarrow 0^{-+}\gamma$
,$0^{++}\rightarrow 1 ^{--}\gamma$  decays take place as $E1$ transition.

Also mixing of $1^{++}$ and $1^{--}$ which
lead to the decays $1^{++}\rightarrow3\gamma$, $1^{--}\rightarrow4\gamma$
$\Gamma(1^{++}\rightarrow3\gamma)=
a_P^2\Gamma(1^{+-}\rightarrow3\gamma)$

Also become possible decays:
\begin{equation}
\label{23}
{}^3 P_1'\rightarrow 3 \gamma,H^0(P^0)+\gamma
\end{equation}
\begin{equation}
\label{A200}
{}^3 S_1'\rightarrow 4\gamma
\end{equation}

Also, Ora-Pauell decay (${}^3 S_1\rightarrow 3g$)
of quarkonium into 3 gluons may be enhanced in the range of soft
gluons,
because due to P-parity violation $1^{--}$ is mixed with $1^{++}$ bound
state which also may decay into three gluons ( for quarkonium decays into gluons
see \cite{N} and references
therein) and maximal in the range of soft gluons (photons).

For $P$-parity violating potential in the system which consist of two
different fermions we obtain the following result in the nonrelativistic
approximation:
\begin{equation}
\label{24}
V(r,\vec{p_1},\vec{p_2})=A+B+C+D
\end{equation}
\begin{equation}
\label{A200}
A=g^V_1g^A_2\frac{1}{2m_1}(Z'(r)\vec{n}[\sigma_1\sigma_2])+2Z(r)\vec{\sigma_2}\vec{p_1}-
i\vec{\sigma_2}\vec{n}Z')
\end{equation}

\begin{equation}
\label{25}
B=g^V_1g^A_2\frac{1}{m_2}(Z(r)\sigma_2\vec{p_2}+\frac{i}{2}\vec{\sigma_2}\vec{n}Z'(r))
\end{equation}
\begin{equation}
\label{26}
C=g^V_2g^A_1\frac{1}{2m_2}(Z'(r)\vec{n}[\sigma_1\sigma_2])-2Z(r)\vec{\sigma_1}\vec{p_2}-
i\vec{\sigma_1}\vec{n}Z'(r))
\end{equation}
\begin{equation}
\label{27}
D=g^V_2g^A_1\frac{1}{m_1}(Z(r)\sigma_1\vec{p_1}-\frac{i}{2}\vec{\sigma_1}\vec{n}Z'(r))
\end{equation}
where $Z^0(r)$ is long range potential considered in \cite{RA}.
Without leptons contribution long range potential has been considered
 also in \cite{1}-\cite{3}.

For fermions with different masses (e.g.$e^-\mu^+,e^-\tau^+,\mu^-\tau^+)$) we
obtain:
\begin{equation}
\label{28}
V(r,\vec{p_1},\vec{p_2})=-\alpha g_Vg_A(A_1+A_2+A_3+A_4)
\end{equation}
where:

\begin{equation}
\label{29}
A_1=\frac{1}{2m_+}(2Z(r)\vec{\sigma_+}\vec{p_+}+i\vec{\sigma_+}\vec{n}Z'(r))
\end{equation}

\begin{equation}
\label{30}
A_2=\frac{1}{2m_-}(-2Z(r)\vec{\sigma_-}\vec{p_-}+i\vec{\sigma_-}\vec{n}Z'(r))
\end{equation}

\begin{equation}
\label{31}
A_3=\frac{1}{2m_-}(Z'(r)\vec{n}[\sigma_+\sigma_-]+2Z(r)\vec{\sigma_+}\vec{p_-}-
iZ'(r)\vec{\sigma_+}\vec{n})
\end{equation}

\begin{equation}
\label{32}
A_4=\frac{1}{2m_+}(Z'(r)\vec{n}[\sigma_-\sigma_+]-2Z(r)\vec{\sigma_-}\vec{p_+}-
iZ'(r)\vec{\sigma_-}\vec{n})
\end{equation}
In case of the fermion-antifermion system with same masses we
have($\vec{p_-}=-\vec{p_+}$):
\begin{equation}
\label{33}
V(r,\vec{p_-})=-\alpha g_Vg_A\frac{1}{2m}(i\vec{n}\vec{S}Z'(r)-iZ(r)\vec{S}\vec{p_-}Z'(r))
\end{equation}
Here  $S_i$ is operator of the total spin:
\begin{equation}
\label{34}
\vec{S}=\frac{1}{2}(\vec{\sigma_1}+\vec{\sigma_2})
\end{equation}

The differential width into 3 photons decays of the ${}^3 S_1$ state (which is
${}^3 S_1$ state with small mixture of ${}^1 P_1$ state) takes the form:

\begin{equation}
\label{35}
\frac{d\Gamma({}^3 S_1'\rightarrow3\gamma)}{d\omega}
\approx a^2_P\frac{d\Gamma({}^3 P_1\rightarrow3\gamma)}{d\omega}+
\frac{d\Gamma({}^3 S_1 \rightarrow3\gamma)}{d\omega}
\end{equation}
where
$\frac{d\Gamma({}^3 P_1\rightarrow3\gamma)}{d\omega}\sim\frac{1}{\omega}$,
whereas $\frac{d\Gamma({}^3 S\rightarrow3\gamma)}{d\omega}\sim\omega log
\frac{m}{\omega}$,
$\omega$ is photon energy, $a_P$ is the measure of the mixing:
\begin{equation}
\label{36}
|{}^3 S_1'>=|{}^3 S_1>+ia_P|{}^3 P_1>
\end{equation}
\begin{equation}
\label{37}
a_P=\frac{<{}^3S_1|V|{}^1P_1>}{E({}^3 S_1)-E({}^1 P_1)}
\end{equation}
Thus at small $\omega$ we have peak instead usual Ora-Pauell
formula. Analogous effect take place in quarkonium (decays into 3
photons,one photon+2 gluons, 3 gluons).
Analogously for above mentioned radiative correction for mixed ${}^3S_1'$ decays we
have:
\begin{equation}
\label{38}
\Gamma({}^3 S_1'\rightarrow {}^1 S_0 \gamma)
\approx
a^2_P\Gamma({}^1 P_1\rightarrow {}^1 S_0 \gamma)+
\Gamma({}^3 S\rightarrow {}^1 S_0 \gamma)
\end{equation}
In bottomonium the wave functions is not Coulomb like, however the
behaviour of the
matrix elements  are expressed through $R(0),R'(0),...$ which may be expressed
through decay widths of the bottomonium(e.g.
$\Gamma(0^{-+}\rightarrow2g)\sim|R_S(0)|^2,
\Gamma(1^{--}\rightarrow3g)\sim|R'(0)|^2$,
see \cite{N} and references therein).

Our results are applicable also to bound states of heavy quarks (several hundred GeV).

For $\tau^+\tau^-$ systems case we obtain using formula (2.5) of the \cite{MM1}
 $a_P=-1.3*10^{-6}$

{\bf $P$-violation in mesoatoms and muonium}

In particular, in $\mu^-p$ atoms the effect
(mixing of S- and P- states with opposite P-parity)
 may  be enhanced in
$\mu^2/m^2_e \approx 4*10^4$($\mu=\frac{m_{\mu}m_p}{m_{\mu}+m_p}$)
times in comparison with hydrogen (See formulas (3.18)-(3.19),(3.22) of \cite{Kh}(see also
references therein)
for $2S_{\frac{1}{2}}-2P_{\frac{1}{2}}$ mixing coefficient in hydrogen).Using esimate of \cite{Kh}
we obtain for range of circular polarization of photon
in $2S_{\frac{1}{2}}$-level decay:
\begin{equation}
\label{39}
P=3.8*10^{-4}\frac{1}{2}(1-4sin\theta_W^2)*(\mu^2/m^2_e)=0.82
\end{equation}
Also may be interesting $P$-violation $\mu$ -mesonic ions where as well
as in \cite{Kh} $P$-violation may be enhanced by large charge of nuclei $Z$.

Also $P$-violation for  $l^+_il^-_j$ atoms may be obtained from
 formulas (3.18)-(3.19),(3.22) of \cite{Kh}. In  particular in case of
$\mu^+e^-$-bound states the effect is the same as in hydrogen.

Analogously may be enhanced CP-violation in mesoatoms in comparison with CP-violation in
hydrogen\cite{Kh} in $\mu^2/m^2_e \approx 4*10^4$ times .

                 {\bf Vector bosons bound states}

 As known \cite{CS},\cite{T} there are the two types of admissible solutions for
  vector bosons bound states in the Coulomb potential:
\begin{equation}
\label{A1}
l=j, \vec{\Phi}=\vec{L}F(r)P_j^m,\phi=0
 \end{equation}
\begin{equation}
\label{A2}
l=j\pm1, \vec{\Phi}=(\vec{L}F_1(r)+i[nL]F_2(r))P_j^m,\phi=G(r)P_j^m
 \end{equation}
(see formulas (35) of ref. \cite{CS})
The solution with $l=j$ lead to the Klein-Gordon equation for
scalar particles, whereas the solution with $l=j\pm1$ lead to the
system of equations which described movement in the potential
which have $r^{-3}$ singularity.As known such singularity lead to
the fall down on the center.If however as we see below  we
consider besides Coulomb field generated by attractive center
the  Higgs field  also generated by the same center (because
attractive particle also obtain its mass via spontaneous symmetry
breaking and consequently interact with Higgs bosons),we obtain
that potential become less singular and in this case fall
down on the center is absent.Also we make the conclusion about
existence of bound states of vector
bosons:$W^+W^-,W^-W^-,W^{++}W^{--},W^{--}W^{--},W^{+}Z^0,H^0Z^0 $etc
which may be produced in $e^+e^-$,$e^-e^-$-collisions.

The part of lagrangian which described vector bosons including its interaction with Higgs bosons
has the following form:
\begin{equation}
\label{A3}
L=-\bar{D}_a\bar{\phi}^bD^a\phi_b+\bar{D}_a \bar{\phi}^bD_b\phi^a+(\phi)^2\bar{\phi}^a\phi_a
\end{equation}
where in accordance with notations \cite{CS} $\phi_a $-is vector bosons ,$\phi$-is Higgs bosons.
Without Higgs bosons we obtain formula (1) of the \cite{CS}.

The field equations are modified due to Higgs bosons presence in comparison with
\cite{CS} by the following way:
\begin{equation}
\label{A3}
(D^aD_a-k^2)\Phi^b=ie\Phi^aH_{ab}-\frac{ie}{2}D_b
 (k^{-2}D_b(H^{ac}F_{ac}))
-D_b(\Phi^a \frac{d}{dx_a}(log(k^2)))
 \end{equation}
where $k^2=g^2\phi^2(r)$.
At large distances $k^2=m^2_W$ and we obtain  equations (18) of
\cite{CS}.
The equation (48) may be rewritten in the following form:
\begin{equation}
\label{A3}
((W+eA_0)^2+\Delta-k^2)\Phi^i=-e\phi A_0'n_i+e\frac{d}{dx_i}
 (k^{-2}((W+eA_0)\vec{n}\vec{\Phi}-\Phi')A_0')-\frac{d}{dx_i}(\vec{n}\vec{\Phi}(log(k^2))')
\end{equation}
where
\begin{equation}
\label{A3}
((W+eA_0)^2+\Delta-k^2)\phi=e\vec{n}\vec{\Phi} A_0'+e(W+eA_0)
 (k^{-2}((W+eA_0)\vec{n}\vec{\Phi}-\phi')A_0')-(W+eA_0)(\vec{n}\vec{\Phi}(log(k^2))'
\end{equation}
If we consider charged vector bosons in the context of Standard
Model (or its various extensions we must add also $Z^0$-bosons
exchanges.It can be made by the following replacements in
equations (49)-(50):
\begin{equation}
\label{A3}
QeA_0(r)\rightarrow QeA_0(r)+(T-2Qsin^2\theta_W)gZ_0(r)
\end{equation}

where $Z_0(r)$ is  $Z$-boson field (radiative corrections
included).

The most singular term in (48) is following:
\begin{equation}
\label{A7}
\sim\frac{ie}{2k^2}D_b(H^{ac}F_{ac})\sim \frac{ieV^2V'}{(m_W+
Cr^{-1}e^{-m_Hr})^2}F_{ac}\sim \frac{e^4r^{-3}}{(m_W+
Cr^{-1}e^{-m_Hr})^2}
 \end{equation}
 If $k^2=m^2_W$ (i.e. constant) we obtain singular potential
 which lead to fall down on the center.
For complete set \footnote{One of this equations is a result of subsidiary condition for
energy level definition the three equations is enough.} of radial equations we obtain the following
result:
\begin{equation}
\label{A17}
\Omega F_1=\frac{2}{r^2}(F_1+j(j+1)F_2)-e\frac{dV}{dr}G-e\frac{d}{dr}(Q\frac{dV}{dr})
+(F_1(log(k^2))'),
 \end{equation}

\begin{equation}
\label{A18}
\Omega F_2=\frac{2}{r^2}F_1-e\frac{1}{r}Q \frac{dV}{dr}-\frac{1}{r}F_1(log(k^2))',
 \end{equation}

\begin{equation}
\label{A19}
\Omega
G=e\frac{dV}{dr}F_1-e(W+eV)\frac{1}{r}Q\frac{dV}{dr}-(W+eV)F_1(log(k^2))',
 \end{equation}

\begin{equation}
\label{A20}
\frac{dlog(k^2)}{dr}F_1+\frac{dF_1}{dr}+\frac{2}{r}F_1+\frac{j(j+1)}{r}F_2+
+e(W+eV)G=-eQ \frac{dV}{dr},
 \end{equation}
where $\Omega$ as in \cite{CS} is Klein-Gordon operator:

\begin{equation}
\label{A21}
\Omega F_2=\frac{d^2}{dr^2}+\frac{2}{r}-\frac{j(j+1)}{r^2}+(W+eV)^2-k^2
 \end{equation}

and $Q$ is define as:

\begin{equation}
\label{A22}
k^2Q=\frac{dG}{dr}-(W+eV)F_1
 \end{equation}
In formulas(48)-(50) $k^2=(m_W+c \frac{\exp{-m_Hr}}{r})^2$.

Case $V=0$ in this equations
correspond to the neutral vector bosons bound states ( $Z^0H^0,Z^0Z^0$ etc.).

At small $r$ we suppose that solutions has the following form:
\begin{equation}
\label{A23}
F_i=A_ir^s
 \end{equation}
\begin{equation}
\label{A24}
G=A_Gr^s
\end{equation}
Substituting (59)-(60) into (53)-(58) we obtain the following system of
linear equations:
\begin{equation}
\label{A25}
-4A_1+A_2(s(s+1)-j(j+1)=0
\end{equation}
\begin{equation}
\label{A26}
A_1(s(s+1)-j(j+1)-2-\alpha(s-1)-2j(j+1)A_2-\alpha(1+\frac{s(s-1)}{c})A_G=0
\end{equation}
\begin{equation}
\label{A27}
(-\alpha+\frac{\alpha^3}{c^2})A_1+(s(s+1)-j(j+1)-2-\frac{s\alpha^2}{c^2})A_G=0
\end{equation}
The determinant of this system must be equal to the zero and we
obtain $s$ as a solutions of this equation.Thus, if $c$ (which characterize
 the presence of Higgs bosons interaction with gauge bosons) is nonzero,
the regular solutions exist.

For $Z^0$-bosons case we obtain:
\begin{equation}
\label{A28}
s=\frac{1\pm \sqrt{1+4j(j+1)+c^2-2}}{2}
\end{equation}
Also if we taking into account repulsive term
$\alpha\frac{(\vec{A})^2}{2m} \sim r^{-4}$ from Klein-Gordon
 equation it also prevent fall down on the center.

Bound states $W^+W^-,W^{++}W^{--},Z^0Z^0,Z^0H^0$ may be produced in
resonance in $ e^+e^-$-collisions:
\begin{equation}
\label{A29}
 e^+e^-\rightarrow W^+W^-,W^{++}W^{--},Z^0Z^0,Z^0H^0
\end{equation}
while $W^-W^-$ -bound states may be produced in
resonance in $ e^-e^-$-collisions:
\begin{equation}
\label{A30}
 e^-e^-\rightarrow W^-W^-,
\end{equation}

It must be noted that bound state may be formed if
$\Gamma<<W$.Thus,the solution $l=j$ is not admissible by this
reason ($\Gamma(W^-)>>W=\frac{1}{4}m_W\alpha^2$.)However for $l=j\pm1$
we expect deep levels ($\Gamma(W^-)<<W=\frac{1}{4}m_W\alpha^2$)
 because we regularized at small distances
singular potential. Besides in case of $W^{++}W^{--}$ bound states
the width of main mode may be $<<$ than $W=\frac{1}{4}m_W\alpha^22^4$
-the binding energy of $l=j$  $W^{++}W^{--}$ bound states.

From the other hand we would like to stress that the deep level
decrees mass of the vector bosons and consequently decrees width
of the vector bosons because it proportional to the mass of the
vector bosons, besides high velocity of the vector bosons in the
bound state also make the width smaller.

 {\bf Vector bosons energy levels in the magnetic field
(Reported on conference "Physics 99" in Yerevan State University 13 September 1999.)}

We choose the magnetic field as $\vec{H}=(0,0,H)$,in this case it is convenient to choose
vector potential in the following form:
\begin{equation}
\label{A31}
A_x=-Hy,A_y=A_z=0
\end{equation}

Substituting $\vec{A}$ into field equation (55)-(56) we obtain the following system of equations
which defined energy levels:
\begin{equation}
\label{A32}
(T-\frac{eH(p_x+eHy)}{k^2}\frac{d}{dy}) \Phi_1=(-ieH+\frac{ieH(p_x+eHy)^2}{k^2})\Phi_2
 \end{equation}

\begin{equation}
\label{A32}
(T+\frac{e^2H^2}{k^2}+\frac{eH(p_x+eHy)}{k^2}\frac{d}{dy}) \Phi_2=(ieH+\frac{ieH}{k^2}\frac{d^2}{dy^2})\Phi_1
 \end{equation}
Here $T= \epsilon^2-(p_x+eHy)^2-p_y^2-p_z^2-k^2$, $p_x$, $p_z$- is
constant.Components $\Phi_{0,3}$ are expressed through $\Phi_{1,2}
\sim \exp{(ip_xx+ip_zz)}\chi_{1,2}(y)$:

\begin{equation}
\label{A33}
T \Phi_3=\frac{-ieH}{k^2}(p_z(p_x+eHy)\Phi_2+ip_z\frac{d}{dy}\Phi_1)
\end{equation}
\begin{equation}
\label{A34}
T \Phi_0=\frac{e^2H \epsilon}{k^2}(-i(p_x+eHy)\Phi_2-\frac{d}{dy}\Phi_1)
\end{equation}

From comparisons of the last two formulas we see that only one of the functions $\Phi_{0,3}$
must be considered as independent.

At large distances
we have $\Phi_{1,2}=A_i\exp{(ip_xx+ip_zz)}\exp{(-\frac{1}{2}eH(y+\frac{p_x}{eH})^2)}$.

After substitution:
\begin{equation}
\label{A35}
 \Phi_i=\exp{(ip_xx+ip_zz)}\exp{(-\frac{1}{2}eH(y+\frac{p_x}{eH})^2)}\chi_{i}
\end{equation}
into equations (68),(69) we obtain:
\begin{equation}
\label{A36}
X''i+(y+\frac{p_x}{eH})A_{ik}X'_k+B_{ik}X_k=0
\end{equation}

where $X_1=\chi_1,X_2=\chi_2-\frac{\omega}{k}\chi_1,
A_{11}=-\omega^2-2eH,A_{12}=0,A_{21}=\omega^2,A_{22}=\omega^2-2eH,
B_{11}=\Omega_0-\omega^2,B_{22}=-\omega^2,B_{21}=\Omega_0+\omega^2-k^2,
B_{22}=\Omega_0+\omega^2,\Omega_0=\epsilon^2-p_z^2-k^2,\omega=\frac{eH}{k}$.

After appropriate diagonalizations we obtain two independent differential equations:
\begin{equation}
\label{A37}
Z_1''-2(y+\frac{p_x}{eH})Z'_1+2n_1Z_1=0,
\end{equation}
\begin{equation}
\label{A38}
Z_2''-2(y+\frac{p_x}{eH})Z'_2+2n_2Z_2=0
\end{equation}
 which solutions are Hermits polinoms if
\begin{equation}
\label{A39}
(P^Ta^{-1}(S^TBS)P)_{11} =-4n_1
\end{equation}
\begin{equation}
\label{A40}
(P^Ta^{-1}(S^TBS)P)_{22} =-4n_2
\end{equation}
where $n_{1,2}$- are integer positive numbers.
In equations (76)(77) $S^TS=I$,$P^TP=I$,$S^TAS=a$ where in matrix $a$
only non-diagonal elements are nonzero.

This conditions (76)(77)  defined energy levels $\epsilon_{1,2}(p^2_z,H,n_{1,2})$
of vector bosons in magnetic field where $\epsilon_{1,2}(p^2_z,H,n_{1,2})$-are
solutions of the equations (76)(77) respectively and must be find numerically.
Substituting $\Phi_{1,2}$ into differential equation ()(which must be solved numerically,
numerical computation are in progress.)
we also obtain energy levels.

At weak  field limit($eH<<k^2$) the term
$\frac{ie}{2}D_b(k^{-2}D_b(H^{ac}F_{ac}))$ may be neglected and we obtain from (68)(69) the following
expression for energy levels:
\begin{equation}
\label{A41}
\epsilon^2 =k^2+p_z^2+2eH(n+\frac{1}{2}\pm \frac{1}{2})
\end{equation}
From (70) we obtain:
\begin{equation}
\label{A42}
\epsilon^2 =k^2+p_z^2+2eH(n+\frac{1}{2})
\end{equation}

New functions $Z_i$ are expressed through $X_i$  by the following way:
\begin{equation}
\label{A43}
Z_i=PSX_i.
\end{equation}

{\bf Magnetism of Electron Gas in the Finite Volume
(Reported on conference "Physics 99" in Yerevan State University 13 September 1999.)}

We also calculate the properties of the electron gas in the finite volume.
We consider two kinds of the finite volumes:
cylinder (with finite length or infinite length) and  in box with sizes (axbxc).
( For energy levels in magnetic field
 and some qualitative discussion about energy levels in magnetic field
inside box has been made in \cite{LL3}).
Due to modification of energy levels also modified
the thermodynamical and chemical potential:
\begin{equation}
\label{85}
\Omega=-kT\sum_mln({1\pm
\exp(\frac{\mu\pm\omega}{kT})})
\end{equation}
\begin{equation}
\label{90}
N=\sum_m\frac{1}{e^(\frac{\omega-\mu}{kT})+1}
\end{equation}
and consequently  magnetization which is expressed through $\Omega$ by the following
way (see e.g. \cite{LL5}):
\begin{equation}
\label{85}
\vec{M}=-\frac{1}{V}(\frac{\partial\Omega}{\partial\vec{H}})_{T,V,\mu}
\end{equation}
If electrons motion is restricted by cylinder
(which described by boundary condition $\psi(R,z)=0$)
we have the following condition which define energy levels:
\begin{equation}
\label{16}
F(-\lambda_{1,2}+l+\frac{1}{2},l+1,\frac{eH}{m_e} R^2)=0
\end{equation}
Here F is hypergeometric function.Using this formula we calculate $M$
 of the  electron gas in the cylinder.Our numerical results
 for energy levels in cylinder are presented on the Fig.1
of the ref.\cite{RA7}.At large $R$ or large $H$ we return to Landau levels.

The ordinary case i.e. particle movement in infinite space $R\rightarrow \infty$
correspond to the:
\begin{equation}
\label{16}
-\lambda+l+\frac{1}{2}=0
\end{equation}
This condition lead to the Landau levels.

If magnetic field is perpendicular to the layer we obtain:
\begin{equation}
\label{60}
E_{m,k}=\sqrt{m^2+|e|H(2m+1+\sigma)+(\frac{\pi k }{a})^2}
\end{equation}
where $m=0,1,2,3...;k=1,2,3...$.

This  formula has been obtained from formula  of the ref.\cite{BLP} on p.148
at $p_z=\frac{\pi k }{a}$

On Fig.2,3 of the ref.\cite{RA7} are presented energy levels for case
of magnetic field which is parallel to the layer.

The magnetization and chemical potential versus $H$ at fixed $R$
(and width in case of layer)
and electrons concentration $n=\frac{N}{V}$ is presented on the Fig.4-Fig.11
of the ref.\cite{RA7}
for cases of the layer (magnetic field is parallel to the layer) and cylinder respectively.

The author express his sincere gratitude to E.B.Prokhorenko for helpful discussions.

\end{document}